\documentstyle[emulateapj,epsfig,apjfonts]{article}
\def\gs{\mathrel{\raise0.35ex\hbox{$\scriptstyle >$}\kern-0.6em 
\lower0.40ex\hbox{{$\scriptstyle \sim$}}}}
\def\ls{\mathrel{\raise0.35ex\hbox{$\scriptstyle <$}\kern-0.6em 
\lower0.40ex\hbox{{$\scriptstyle \sim$}}}}

\newcommand{\Lsun}{\mbox{$L_\odot$}}

\newcommand{\mum}{$\,\mu$m}

  \def\itm#1 {\vskip10pt \noindent \square\ {\bf #1} }
  \def\square {\hbox{\vrule width5pt height5pt}}

  \def\cle      {{$_ <\atop{^\sim}$}}
  \def\cge      {{$_ >\atop{^\sim}$}}

  \def\Lsun     {{$L_{\odot}$} }

  \def\deg      {{\ifmmode^\circ\else$^\circ$\fi} } 
  \def\arcm    {{\ifmmode {'\ }\else$'     $\fi} } 
  \def\arcs    {{\ifmmode{''\ }\else$''    $\fi} } 
\lefthead{Chapman et al.}
\righthead{The bright sub-mm galaxy population}

\begin{document}

\title{The nature of the bright submillimeter galaxy population:
a radio pre-selected sample with $I>25$} 

\author{S.\,C.\ Chapman,$\!$\altaffilmark{1} E.\ Richards,$\!$\altaffilmark{2}
G.\,F.\ Lewis,$\!$\altaffilmark{3}
G.\ Wilson,$\!$\altaffilmark{4,5} A.\ Barger$\!$\altaffilmark{4,6}
}
\affil{$^1$Observatories of the Carnegie Institution of Washington, Pasadena,
Ca 91101,~~U.S.A.}
\affil{$^2$Department of Physics, University of Alabama,
           Huntsville, AL 35899, U.S.A.}
\affil{$^3$Anglo-Australian Observatory, P.O. Box 296, Epping,
NSW 1710,~~Australia}
\affil{$^4$Institute for Astronomy, University of Hawaii, 2680 Woodlawn Dr,
Honolulu, HI 9682,~~~U.S.A.}
\affil{$^5$Physics Department, Brown University, Providence, RI 02912,~~~U.S.A.}
\affil{$^6$University of Wisconsin, 475 N Charter Street, Madison, WI 53706,~~~U.S.A.}
        


\begin{abstract}
Deep submillimetre surveys have
  successfully detected distant, star-forming galaxies, enshrouded in vast
  quantities of dust and which emit most of their energy at far infrared
  wavelengths. These luminous galaxies are an important
  constituent of the Universal star-formation history, and any complete
  model of galaxy evolution must account for their existence. Although
  these sources have been tentatively identified with very faint and
  sometimes very red optical counterparts, their poorly constrained redshift
  distribution has made their interpretation unclear. In particular, it was
  not understood if these galaxies had been missed in previous
  surveys, or if they constituted a truly new class of objects,
  undetectable at other wavelengths.  By utilizing a radio selection
  technique, we have isolated a 
sample of 20 sub-mm objects representative
  of the 850\mum\ population brighter than 5\,mJy with $z$\cle 3. 
We show that these
  galaxies are so heavily dust obscured that they remain essentially
  'invisible' to ultraviolet selection. 
Furthermore, relying on the radio-submillimeter flux
density ratio, we estimate their redshift
  distribution, finding a median of two. These results are inconsistent
  with the existence of a very high redshift ($z>4$) population of primeval
  galaxies (L$_{bol} > 10^{12}$\,h$^{-2}$\Lsun)
 contributing substantially to the sub-mm counts.
While not a substitute for the thorough followup of blank field sub-mm
surveys, our results do shed light on a substantial portion of the
luminous sub-mm population with $z$\cle 3.
\end{abstract}


%
%
%
\section{Introduction}

The extragalactic far-infrared background light is
  believed to be composed of the integrated thermal starlight
and non-thermal AGN radiation, reradiated by dust
  within star-forming galaxies over the entire history of galaxy formation.
  The energy density of this infrared background is approximately the same
  as found in the optical suggesting that at least half of the Universal
  star-formation history remains hidden from optical view (Puget et al.~1996). 
This diffuse background was first resolved into discrete sources by the
  Sub-millimetre Common User Bolometer Array (SCUBA -- Holland et al.~1999)  
on the James
  Clerk Maxwell Telescope (JCMT) by a number of groups (Smail et al.~1997,
Hughes et al.~1998, Barger et al.~1998, Eales et al.~1999).
  
  Although a large number of deep SCUBA surveys has led to a better
estimate of the 850 micron galaxy surface density, our understanding of
the nature of the sub-mm population remains limited. The principal
obstacle is obtaining reliable counterparts of these sub-mm sources
at other wavelengths, a problem exacerbated by both the
coarseness of the JCMT resolution (15 arcsec at 850 microns) and the
inherent faintness of suspected optical counterparts (Smail et al.~1999). 
It is still
unclear if the sub-mm selected sources are related to known populations,
such as high redshift quasars (Hughes et al.~1997, McMahon et al.~1999) 
or Lyman-break galaxies 
(Chapman et al.~2000a), or
constitute a truly new class of objects. With the exception of several
isolated objects, few reliable identifications have been made (e.g.
Ivison et al.~1998, Frayer et al.~1998). 
Thus the redshift distribution has remained largely unconstrained
over a vast range, with the possibility that many sources lay at extreme
distances ($z > 5$).
  
  One technique, which has shown some promise in identifying sub-mm
sources, is radio continuum followup. Because galaxies and the
inter-galactic medium are transparent at centimeter wavelengths, radio
emission is unhindered by intervening gas and dust. Ubiquitous in local
star-forming galaxies, radio emission also correlates very strongly with
the far-infrared emission in star-forming galaxies (Helou et al.~1986).  
Moreover, the
high resolution provided by radio interferometers can provide a surrogate
for the poor sub-mm angular resolution and astrometric uncertainties.  

  Given the difficulties of obtaining secure optical identifications and
spectroscopic redshifts, the radio observations provide another clue to the nature of
sub-mm sources. Via the empirically observed far-infrared to radio
correlation in local star-forming galaxies, one can use the observed ratio
of sub-mm to radio continuum flux density to estimate a redshift.  As the
k-corrections (corrections based on the redshifted spectral energy
distribution - SED) of the radio and sub-mm flux densities are opposite in
slope, the ratio of radio to sub-mm flux density is quite sensitive to
redshift (Carilli \& Yun 1999). 

Barger, Cowie \&
Richards (2000 -- hereafter BCR) first attempted to use a radio
selected sample to target a number of optically faint microJansky radio
sources with near-infrared magnitude, $K > 20.5$. Using SCUBA to a 3$\sigma$
RMS limiting flux density of 6\,mJy at 850 microns, they detected 5 out of
15 radio sources, while in the process demonstrating that none of the
optically brighter radio sources ($K < 20.5$) 
were detected in the sub-mm. The surface
density of these few bright radio selected sub-mm sources closely matched
that from blank field surveys, indicating a close correspondence between
the optically faint radio population and bright sub-mm sources. Other
pointed SCUBA studies of known high-$z$ populations such as $z\sim3$ Lyman-break
galaxies (LBGs -- Chapman et al.~2000a) 
and radio loud AGN (Archibald et al.~2000) 
have revealed few SCUBA
detections, and nowhere near the surface density of blank field sub-mm
sources.  

 We have refined the selection criterion to those microJansky
radio sources with an optical magnitude, $I > 25$, based on the 
clear bi-modal break in the optical properties of microJy radio sources
(Richards et al.~1999). We have applied this technique to a 
sample in the region surveyed by Richards (2000) centered on
the Hubble Deep Field. We have selected a total of 47 radio sources in our
study, 20 previously observed, which meet our criterion. Our followup
SCUBA {\it photometry} 
observations of 27 radio selected objects demonstrate this to be a
highly successful technique. We now detect $\sim$50\% of the new 27 object
sample above 4.5\,mJy at 850 microns, with an overall success rate of 20
out of 47 objects observed. Our increased detection success over BCR is
likely to be a result of our slightly deeper survey coupled with the
stricter selection criterion.  
We are thus able to
uncover bright sub-mm sources
using SCUBA at the rate of
one source per hour on the JCMT, greater than
an order of magnitude
more rapid than mapping a random patch of sky.
Our new survey represents a sub-mm mapping of a $\sim$100 arcmin$^2$ effective
region in less than 16 hours,
sensitive to sources S$_{850}>5$\,mJy and $z$\cle 3.

\section{Source Selection and Sub-mm Observations}

 The HDF region has previously been imaged at 1.4\,GHz
using the Very Large Array (VLA) radio telescope by Richards (2000) 
to a completeness of 40\,$\mu$Jy. We aligned a deep $I$-band optical
image described in Barger et al. (1999), with the VLA FK5 astrometric
frame by using 102 of 333 radio sources which lie within the 
30\arcm$\times$30\arcm\
optical field of view.  After alignment, 60 radio sources were determined
not to have any optical identifications within 2 arcsec of the radio
position brighter than $I<25$. Of these, 17 have previously been described
in Richards et al. (1999). Several of these were followed up as part
of the sub-mm study of BCR. Of the increased sample of 60 optically faint
radio objects, we chose to concentrate on the 40 which had previously not
been observed by BCR or Hughes et al.~(1998). 
Time constaints allowed us to observe only 27
of these, although these were selected at random with no further selection
bias.  

 We observed each of the radio sources using the JCMT/SCUBA at
850/450 microns in photometry mode for an effective integration time
between 600s and 2000s. The secondary was chopped at 7.8125 Hz, using a
chop throw ($\sim$50\arcsec) and direction chosen to keep the source in a bolometer
throughout the observation. Pointing was checked before and after the
observation on blazars and sky-dips were performed to measure the
atmospheric opacity directly. The RMS pointing errors were below 2 arcsec,
while the average atmospheric zenith opacities at 450 and 850 microns were
1.7 and 0.24 respectively. The data were reduced using the Starlink
package SURF (Scuba User Reduction Facility -- Jenness et al.~1998), 
and our own reduction
routines to implement the three bolometer chopping mode. Spikes were first
carefully rejected from the data, followed by correction for atmospheric
opacity and sky subtraction using the median of all the array pixels,
except for obviously bad pixels and the source pixels. The data were then
calibrated against standard planetary and compact HII region sources,
observed during the same nights.  

Additionally, we reanalysed the SCUBA data from BCR and found their
source `3' to have a 3$\sigma$ detection (5.3$\pm$1.7\,mJy), which
we include in our present sample.
We also include the sub-mm source {\it HDF850.2} 
from the Hughes et al.~1998
study, having a radio source counterpart S$_{1.4}>$40$\mu$Jy and $I>25$.

\section{Analysis and Discussion} 

The crucial data available to us from our technique are the optical
properties and redshift estimates for the sub-mm sources, which we present
in Table~1.  Our results assume a
$\Lambda=0.0$, $\Omega=1.0$, H$_0$=65\,km/s/Mpc cosmology.  
Redshift estimates can
be obtained from the sub-mm/radio flux ratios (Carilli \& Yun 2000). 
All of our sub-mm sources fall roughly in the redshift
range $z=1-3$ with a median redshift for the sample of $z=1.9$, consistent
with previous results from BCR. 
The sensitivity of our radio survey to star-forming galaxies
with radio luminosities fainter than $10^{24}$\,W/Hz diminishes
quickly past $z\sim3$, and hence biases our sub-mm survey.
An independent check on the sub-mm/radio redshift estimates can be obtained
through the 450\mum/850\mum\ ratio (e.g. Hughes et al.~1998).
Subject to unknown dust temperature, we obtain 
an estimate of T$_{\rm d}$/(1+$z$), which we list in Table~1 for 
T$_{\rm d}$=45\,K for consistency with the ultra-luminous infrared galaxy,
Arp\,220. 
Raising or lowering the adopted dust temperature 
has the effect of a corresponding systematic raising and
lowering of both our redshift estimates (Blain 1999).

        In order to calculate the density of sub-mm
sources on the sky as found in our radio pre-selection
survey, we need to determine the effective area covered by
our study. This is given simply by the overlap regions
between the optical and radio images. However, a
further complication arises from the non-uniform
sensitivity of the radio image which serves
to decrease the visibility area. This issue is
discussed in Richards (2000) and we use the same
method for determining the source count.         

Our pre-selected sample has already saturated the bright ($>5$\,mJy) sub-mm
counts (Fig.~1), 
and there are not likely to be many additional bright, high
redshift sub-mm sources in our survey region. BCR found that 2 additional
sub-mm sources without radio counterparts were detected in their survey
area, indicating 75\% of the bright sub-mm sources are typically recovered
through such radio pre-selection.  Since we performed photometry on the
sources, we have no means of estimating this extra population.  
By accounting for such an additional 25\% high redshift population 
our sample is in agreement with
analyses of the redshifts for
lensed sub-mm sources (Barger et al.~1999, Smail et al.~2000, 
Blain et al.~1999b). 
The percentage of sub-mm sources missed by our pre-selection technique will
depend field to field on the high-$z$ clustering of sub-mm luminous sources.
A lensed sub-mm survey (Smail et al.~2000) 
detects a similarly large fraction of their 
bright sub-mm sources in the radio.
Deeper blank field sub-mm surveys 
(e.g. Eales et al.~2000 -- S$_{850\mu m}$\cge 3\,mJy)
detect $\sim$1/3 of their sources in the radio. This is roughly in agreement
with our results which use deeper radio maps and brighter sub-mm limits.
While not a substitute for the thorough (and difficult) 
followup of blank field sub-mm
surveys, our results do shed light on a substantial portion of the
luminous sub-mm population with $z$\cle 3.

 At the faint end of the counts, our
pre-selected sources appear to fall short of full source counts recovery,
and the interesting question becomes what is the nature of the 850-micron
sources at flux densities $<$5\,mJy? Although we are only sensitive to
sources with S$_{850}$\cge 5\,mJy, we can deduce important properties about the
fainter sub-mm population. Averaging our sub-mm undetected
sample (inverse variance weighted) reveals a mean flux of
S$_{850}=0.8\pm0.3$\,mJy, suggesting that 
many of these $\sim$50\%
of our $I>25$ and S$_{1.4}>40$\,$\mu$Jy radio sample are 
still fairly luminous sub-mm sources. They likely form a continuous
distribution with the S$_{850}$\cge 5\,mJy sample,
lie at similar or lower redshifts (Table~1), and
comprise roughly 10\% of the blank field SCUBA source counts
from 1--5\,mJy.  This
leaves a large portion of sub-mm sources fainter than 5\,mJy that are not
subsumed in our present $I>25$ radio sample.  

 The first possibility is
that these fainter sub-mm sources are largely contained in our present
radio surveys, but are actually optically bright ($I<25$).  Bright LBGs are
known to emit at the 1--2\,mJy 
level (Chapman et al.~2000a, Peacock et al.~2000), 
so they must contribute some
fraction of the missing sources.  This is consistent with the high SFR
deduced by Steidel et al.~1999, 
who apply a large dust correction to
their results.  Our Arp\,220 SED model suggests that these sources would be
detected in our present radio survey almost out to $z=3$.  However, the
HDF-SCUBA results (Hughes et al.~1998, Peacock et al.~2000) 
also show directly that a significant fraction of
the 2\,mJy sources are not associated with bright LBGs.  Although deeper
radio observations with optically faint counterparts may quickly recover
this population, it also remains a possibility that these sources
represent high redshift ($z>4$) protogalaxies with 
L$_{\rm bol} < 10^{12}$\Lsun, which
would remain undetected in the radio to significantly deeper flux limits.  
So by pushing to fainter radio limits, it is rather unclear what types of
objects might be pre-selected.  

 An additional concern with the sub-mm
source population is that massive AGN may be heating the dust rather than
star formation. We can be reasonably certain that our radio selected
sub-mm sources are not driven primarily by AGN for two reasons. Firstly,
the sources are spatially resolved with a median of about 2\arcsec\ 
in the radio using the Merlin
interferometer at a resolution of 0.2\arcsec\ 
(Richards 2000, Muxlow et al.~2000), 
corresponding to $\sim$1\,kpc at
$z=2$ for our adopted cosmology. If the radio and associated sub-mm emission
were emanating from such a compact active nucleus (AGN), it would appear
unresolved even at this fine resolution.  Secondly, the sub-mm sources
have recently been shown to have little or no overlap with X--ray sources
as observed with the Chandra satellite (e.g. Fabian et al.~2000, 
Hornschemeier et al.~2000,
Barger et al.~2001). As scenarios which
would obscure even the X--ray emission are improbable, the implication is
that most bright sub-mm sources are in fact driven by star formation.  

 Assuming then that our radio-selected sources are driven primarily by
star formation, it is appropriate to use our data to estimate a contribution
to the comoving star formation rate density (SFRD).  In Fig.~2, we
integrate 
over our measured counts in three redshift bins,
and divide by the effective volume of
our detected sources, 
to represent the sub-mm SFRD as a function of
redshift. 
Since we have assumed the far-IR/radio relation in the redshift
estimates, our SFRD estimates can be derived from either wavelength in a
manner similar to BCR. We plot our new 
points as solid squares (sub-mm detected
objects) and a solid hexagon (sub-mm undetected objects).  The very high
redshift open square ($z>3$) represents the Hubble Flanking Fields
sub-mm sources undetected
in the radio from BCR and Borys et al.~2001.  
We then compare optically selected sources (open triangles) 
at redshifts $1<z<4$ (Connolly et al.~1997, 
Steidel et al.~1999). 
We apply a dust correction to the optical points for $z>1$, using
the Steidel et al.~(1999) prescription (factor $\sim$5 for $z>2$). This
is roughly in accord with the expected correction from the Chapman et
al.~(2000b) sub-mm measurement of 33 LBGs with large expected 
star formation rates.  Lower redshift radio
selected sources with bright optical counterparts, as analysed by
Haarsma et al.~(2000) are also plotted 
(red open circles) requiring no correction factor for dust obscuration.

 Sub-mm sources fainter than 5\,mJy likely
begin to merge with optically selected samples (e.g. Adelberger \& Steidel
2000).  However, our $I>25$ radio selected population
is truly an orthogonal population to those
discovered in optical surveys, even for S$_{850}$\cle 5\,mJy.  
We can then confidently sum the optically selected (but uncorrected for dust)
($I<25$) and sub-mm ($I>25$) points from $z=$1--4 to arrive at a 
conservative lower limit to the total SFRD of
all presently known objects (plotted as large stars with arrows).
This lower limit can be compared with the dust corrected optical points,
which is still a rather uncertain procedure.

We have therefore
recovered the majority of the bright ($>5$\,mJy) sub-mm sources with a
statistically significant sample, selected based on the microJy radio
emission with extremely faint optical counterparts. We can state with some
assurance that the bulk of bright sub-mm selected sources represent a
highly dust obscured star forming population which would be very difficult
to identify in optically based surveys. Our redshift analysis has also
demonstrated that this is not because the sub-mm sources are at very high
redshifts ($z>>3$). 
These sources may represent the epoch in which the most
massive spheroid galaxies were being formed through merging fragments in
cluster environments. Indeed, the recent identification of prodigious
sub-mm emission with highly over-dense cluster 
cores at $z\sim3$ and $z\sim3.8$ (Chapman et al.~2000c and Ivison et
al.~2000 respectively) suggests
that the most luminous members of our sub-mm source sample may highlight
similar such regions.  Pushing our study to fainter sub-mm and radio flux
limits will facilitate our understanding of the transition and overlap
between these ultra-luminous high$-z$ star formers (which may evolve into
the most massive spheroids in the present epoch) and the less massive
galaxies selected in the optical through the Lyman-break technique 
(Steidel et al.~1999).  

\acknowledgments
We thank the  staff of the JCMT
for their assistance with the SCUBA observations.
The James Clerk Maxwell Telescope is operated by
The Joint Astronomy Centre on behalf of the Particle Physics and
Astronomy Research Council of the United Kingdom, the Netherlands
Organization for Scientific Research, and the National Research
Council of Canada.

%
%
\begin{figure}
\begin{center}
\epsfig{file=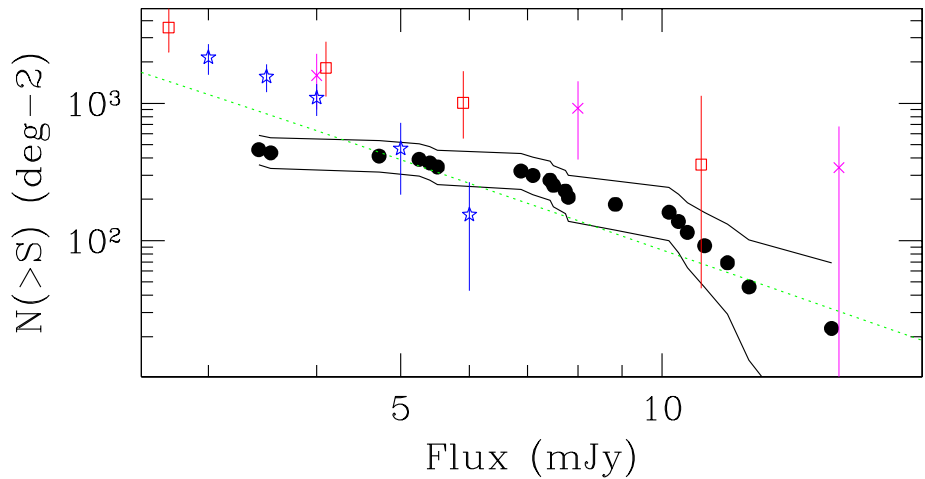,height=7.25cm,angle=0}
\end{center}
\figcaption[chapman.fig1]{
The integrated source counts estimated from our
optically faint radio source sample (solid dots).
The jagged solid lines are 1$\sigma$ uncertainties.
Also plotted are the
fit to blank field sub-mm survey counts from BCR (solid line),
the counts of Eales et al.~(2000 -- stars), and the
lensing amplified counts of Blain et al.~(1999 -- crosses) and
Chapman et al.~(2001 -- squares).
Our radio selected sources saturate the bright sub-mm counts, and
there are not likely to be many additional bright, high redshift sub-mm
sources in our survey region.
}
\label{f1}
\end{figure}

%
%
\begin{figure}
\begin{center}
\epsfig{file=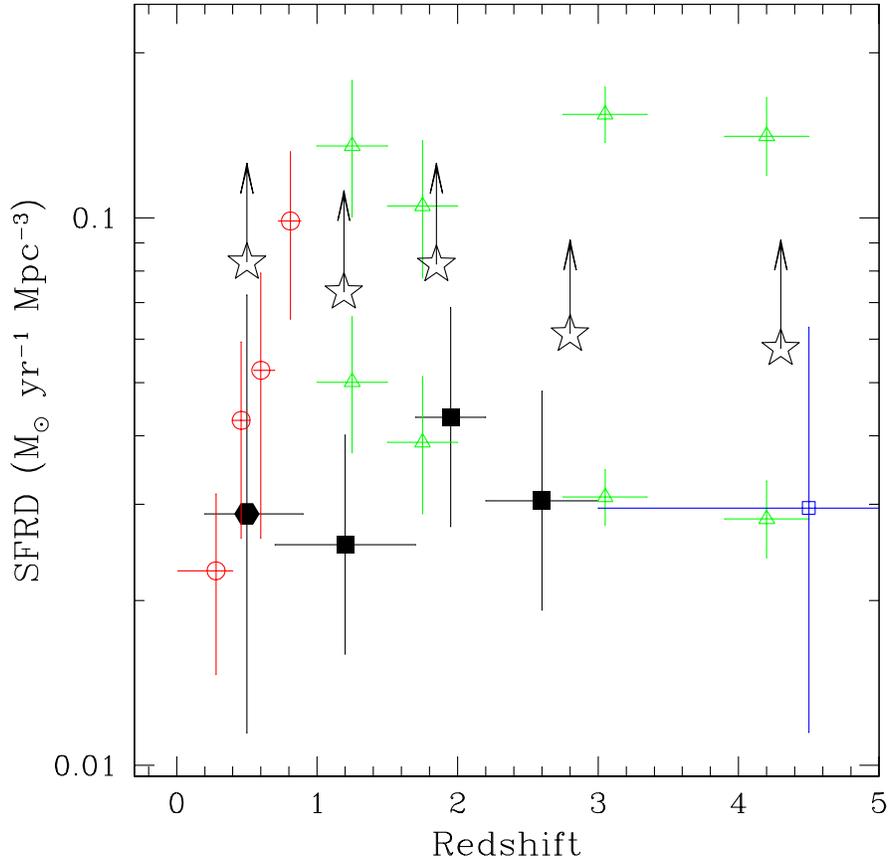,height=12.25cm,angle=0}
\end{center}
\figcaption[chapman.fig2]{
The star formation rate
density as a function of redshift.
Measurements are obtained by integrating over our the source counts, and
dividing by the effective volume of the detected sources
(assuming a $\Lambda$=0, $\Omega$=1, $H_0$=65\,km/s/Mpc cosmology).
Our new sub-mm
  points are represented as solid squares (sub-mm detected sources) and
a solid hexagon (sub-mm undetected sources),
while the very high redshift point ($z>3$ - open blue square)
  represents the sub-mm sources undetected in the radio from BCR.
Compare these points with optically-selected samples with and without
dust corrections (see text) over the
range $1<z<4$, represented as green open triangles (Connelly et al.~1997,
Steidel et al.~1999), and radio-selected
samples (Haarsma et al.~2000)
with confirmed optical counterparts ($I<25$), plotted as
red open circles.
The optical and sub-mm measured points thus plotted
represent essentially orthogonal galaxy populations
($I>25$ and $I<25$),
and we can confidently sum the optically selected (but uncorrected for dust)
($I<25$) and sub-mm ($I>25$) points from $z=$1--4 to arrive at a
conservative lower limit to the total SFRD of
all presently known objects (plotted as large stars with arrows).
This lower limit can be compared with the dust corrected optical points,
still a rather uncertain procedure.}
\label{f2}
\end{figure}

%
%
\begin{deluxetable}{lrrcc}
\tablewidth{280pt}
\scriptsize
\tablenum{1}
\label{table-1}
\tablecaption{\sc \small The radio and sub-mm properties of our new
SCUBA observed sources, along with the redshift estimate assuming the SED
of Arp220.  Note that all sources have $I>25$, the 5$\sigma$ completeness
limit of our optical imagery.\label{tab1}}
\tablehead{
\colhead{source} &  \colhead{$S_{1.4{\rm GHz}}$}
 & \colhead{$S_{850\mu{\rm m}}$} & \colhead{$z_{CY}^{\rm a}$}
 & \colhead{$z_{450/850}^{\rm c}$}\\
\colhead{} & \colhead{($\mu$Jy)} & \colhead{(mJy)} & \colhead{} &
 \colhead{}  }
\startdata
1&     143$\pm$13.2  &    8.3$\pm$2.8 & 2.2$^{+1.2}_{-0.8}$ & n/a\\
2&    98.7$\pm$10.3  &    5.5$\pm$1.8 & 2.0$^{+1.1}_{-0.8}$ & $>1.7$\\
3&    58.4$\pm$9.00   &    8.8$\pm$2.1 & 2.9$^{+1.4}_{-1.0}$ & $>1.1$\\
4&     212$\pm$13.7  &    5.4$\pm$1.9 & 1.3$^{+0.9}_{-0.6}$ & $>1.0$\\
5&     262$\pm$17.1  &    6.9$\pm$2.0 & 1.5$^{+0.9}_{-0.6}$ & $>1.9$\\
6&    74.4$\pm$9.00  &   11.6$\pm$3.5 & 3.0$^{+1.5}_{-1.0}$ & n/a\\
7&    78.8$\pm$9.10  &   10.4$\pm$3.4 & 2.9$^{+1.4}_{-1.0}$ & n/a\\
8&     264$\pm$18.4  &    3.5$\pm$1.2 & 1.0$^{+0.8}_{-0.5}$ & $>0.5$\\
9&     660$\pm$35.8   &   4.7$\pm$1.6 & 0.6$^{+0.4}_{-0.3}$ & $>0.5$\\
10&    170$\pm$12.8  &   10.2$\pm$2.7 & 2.0$^{+1.2}_{-0.8}$ & $>1.3$\\
11&    551$\pm$30.6  &    7.4$\pm$2.2 & 1.0$^{+0.8}_{-0.5}$ & $>0.4$\\
12&    132$\pm$10.1  &    7.7$\pm$2.4 & 2.0$^{+1.2}_{-0.8}$ & n/a\\
13&    595$\pm$30.9  &   15.7$\pm$2.4 & 1.4$^{+0.9}_{-0.6}$ & $>0.9$\\
14&    378$\pm$25.9 &      -0.6$\pm$2.5 & $<$1.0 & \\
15&    117$\pm$11.9  &     0.7$\pm$2.1  & $<$1.6 & \\                    
16&    193$\pm$14.7  &     -1.2$\pm$2.1 & $<$1.2 & \\
17&    118$\pm$10.6  &     2.3$\pm$1.6  & $<$1.4 & \\
18&    354$\pm$22.2  &     2.0$\pm$1.9 & $<$0.9 & \\
19&   55.4$\pm$9.10  &    2.1$\pm$2.4  & $<$2.2 & \\
20&   80.6$\pm$9.30  &    2.1$\pm$2.2  & $<$1.9 & \\
21&    111$\pm$11.4  &     -0.9$\pm$2.2 & $<$1.6 & \\
22&    159$\pm$11.7  &     3.2$\pm$2.3  & $<$1.4 & \\
23&    202$\pm$15.4  &     1.5$\pm$2.2 & $<$1.2 & \\
24&    307$\pm$17.4  &     -0.7$\pm$2.2 & $<$1.0 & \\
25&    506$\pm$30.2  &     -4.6$\pm$3.0 & $<$0.9 & \\
26&    708$\pm$36.9  &     3.2$\pm$1.9  & $<$0.6 & \\
27&   6580$\pm$330   &    -0.5$\pm$3.7  & $<$0.02 & \\

\enddata
\vspace*{-0.5cm}
\tablerefs{{}\\
a) Redshift estimates from the sub-mm/radio index using the
Carilli \& Yun (2000) indicator.\\
b) An independent check on the sub-mm/radio redshift estimates
through the 450\mum/850\mum\ limit (e.g. Hughes et al.~1998)
and a dust temperature T$_{\rm d}$=45\,K, consistent with Arp220.
The 450\mum\ measurement is often not sufficient to provide a
useful limit (listed `n/a').
}
\end{deluxetable}

\end{document}